\documentstyle[fleqn,12pt,epsf]{article}
\newcommand{\Kbar}{\not{\!K}}
\newcommand{\Hbar}{\not{\!H}}

\newcommand{\Pbar}{\not{\!P}}

\newcommand{\be}{\begin{equation}}
\newcommand{\ee}{\end{equation}}
\newcommand{\ba}{\begin{eqnarray}}
\newcommand{\ea}{\end{eqnarray}}

\newcommand{\tauvec}{\mbox{\boldmath $\tau$}}

\newcommand{\nsigma}{\mbox{\boldmath $\sigma$}}
\newcommand{\npi}{\mbox{\boldmath $\pi$}}
\newcommand{\ngamma}{\mbox{\boldmath $\gamma$}}

\newcommand{\nh}{{\bf      h}}
\newcommand{\nj}{{\bf      j}}
\newcommand{\nk}{{\bf      k}}
\newcommand{\np}{{\bf      p}}       
\newcommand{\nq}{{\bf      q}}

\newcommand{\nA}{{\bf      A}}       
\newcommand{\nB}{{\bf      B}}

\newcommand{\nT}{{\bf      T}}

\begin{document}
\begin{titlepage}
\mbox{} 
\vspace*{2.5\fill} 
{\Large\bf 
\begin{center}

Delta-isobar relativistic meson exchange 
currents in quasielastic electron scattering

\end{center}
} 
\vspace{1\fill} 
\begin{center}
{\large 
J.E. Amaro$    ^{1}$, 
M.B. Barbaro$  ^{2,3}$, 
J.A. Caballero$^{3}$, 
T.W. Donnelly$ ^{4}$ and 
A. Molinari$   ^{2}$
}
\end{center}
\begin{small}
\begin{center}
$^{1}${\sl 
Departamento de F\'\i sica Moderna,
Universidad de Granada, 
E-18071 Granada, SPAIN 
}\\[2mm]
$^{2}${\sl 
Dipartimento di Fisica Teorica,
Universit\`a di Torino and
INFN, Sezione di Torino \\
Via P. Giuria 1, 10125 Torino, ITALY 
}\\[2mm]
$^{3}${\sl 
Departamento de F\'\i sica At\'omica, Molecular y Nuclear \\ 
Universidad de Sevilla, Apdo. 1065, E-41080 Sevilla, SPAIN 
}\\[2mm]
$^{4}${\sl 
Center for Theoretical Physics, Laboratory for Nuclear Science 
and Department of Physics\\
Massachusetts Institute of Technology,
Cambridge, MA 02139, USA 
}
\end{center}
\end{small}

\kern 1. cm \hrule \kern 3mm 

\begin{small}
\noindent
{\bf Abstract} 
\vspace{3mm} 

We study the role of the $\Delta$-isobar current on the response
functions for high energy inclusive quasielastic electron scattering
from nuclei.  We consider a general Lagrangian which is compatible
with contact invariance and perform a fully relativistic calculation
in first-order perturbation theory for one-particle emission.  The
dependence of the responses upon off-shell parametrizations is
analyzed and found to be mild.  A discussion of scaling behaviour and
a comparison with various non-relativistic approaches are also
presented.

\kern 2mm 

\noindent
{\em PACS:}\  25.30.Fj, 25.30.Rw, 14.20.Gk, 24.10.Jv, 24.10.Cn

\noindent
{\em Keywords:}\ Nuclear reactions; Inclusive electron scattering;
Meson-exchange currents; Relativistic Fermi Gas; $\Delta$-isobar current.

\end{small}
\kern 2mm \hrule \kern 1cm
\end{titlepage}

\section{Introduction}

In recent work~\cite{Ama02a,Ama02b,Ama02c} 
we have studied the role of pionic
correlations for electron-nucleus scattering in the kinematical domain
of the quasielastic peak, focussing our attention on the 1p-1h sector. The
calculation has been carried out on the basis of a Relativistic Fermi
Gas (RFG) model in which an exact relativistic treatment of the problem
can be accomplished. In the domain of perturbation theory, all
diagrams including one pionic line were included --- in particular the
meson-exchange currents (MEC) of contact and pion-in-flight types ---
so that the corresponding current was automatically gauge invariant.

In \cite{Ama02a,Ama02b} the diagrams involving the excitation of a
virtual $\Delta$-resonance, represented in fig.~1, were
neglected. Although they do not affect the gauge invariance of the
theory, since they are associated with a conserved current, it is
well-known that such contributions do modify the nuclear response
functions, especially in the transverse channel.  The aim of this
paper is to extend the model of refs.~\cite{Ama02a,Ama02b} to
include the $\Delta$-induced meson-exchange-currents of fig.~1,
in order to have a complete understanding of pionic effects in the
quasielastic peak (QEP) as far as the 1p-1h channel is concerned. In
contrast with non-relativistic calculations~\cite{Alb90} our results
do not involve any expansions in energy-momentum/$M$, where $M$ is
a typical baryonic mass, and can therefore be applied when studying
the response at high momentum and energy transfers.

The present relativistic treatment of the $\Delta$ current 
allows one to study
several aspects of $\Delta$ electroexcitation
which are of special interest theoretically.  In particular, 
in the
effective Lagrangian approach to electroexcitation of the $\Delta$
resonance it is known that there is freedom at the electromagnetic
(EM) vertex due to the
off-shell behavior of this resonance~\cite{Ben89}. In contrast to
earlier approaches to the $\Delta$-exchange current in a relativistic
model~\cite{Dekker}, here we consider a more general $\gamma N\Delta$
interaction Lagrangian.  First of all, current conservation restricts
the form of the vertex to a superposition of three covariants, for
which one can use any of the choices described for instance by Jones
and Scadron~\cite{Jon73}. One possible choice is the familiar set of
magnetic dipole, electric quadrupole, and Coulomb quadrupole
multipoles, used in the pion electroproduction analyses 
of~\cite{Noz90,Van95}. In this work we use instead the standard 
``normal parity'' set, analogous to the Dirac-Pauli decomposition of 
the nucleon form factor~\cite{Gar93}. 
The corresponding term in the
traditional chiral Lagrangian of Peccei~\cite{Pec69} is just a
particular case of the normal parity set with two of the three terms
equal to zero.

Going a step beyond the Jones-Scadron vertex, the off-shell
propagation of massive vector fields with spin $S>\frac12$ has been
exhaustively discussed in the literature~\cite{Ben89,Hab98,Pas99}. In
particular, a spin $\frac32$ field generates contributions involving
the $S=\frac12$ sector of this field in the effective amplitudes.  In
the case of the $\Delta$, this is obtained by constructing the most
general vertex which is invariant under a special contact symmetry of
the free Lagrangian (see \cite{Ben89} for details).  The invariance of
the $\gamma N\Delta$ and $\pi N \Delta$ vertices under this ``point
transformation'' of the field requires the introduction of additional
parameters in the Lagrangian. Attempts have been made to fix some of
these ``off-shell'' parameters by fitting the pion electro- and
photo-production data~\cite{Dav91} and, more recently, Compton
scattering data from the nucleon~\cite{Pas95}.  In this work we
analyze the impact of the off-shellness nature of the $\Delta$ in the
EM responses for high momentum transfers by comparing the results
obtained with different sets of parameters fitted to nucleon data.  In
this context we study, in particular, the differences with the
traditional Peccei Lagrangian approach, which corresponds to specific
choices of the off-shell parameters and coupling constants.  Recently
attempts have been made to design new $\Delta$ interaction lagrangians
``consistent'' with the number of spin degrees of freedom of the
$\Delta$ from a rigorously field theoretical point of
view~\cite{Pas99,Pas98,Pas01}. However analyses of pion
photoproduction with these interactions have not been performed yet,
and these studies should be done before attempting to implement them
into the MEC operators.

\begin{figure}[t]
\begin{center}
\leavevmode
\def\epsfsize#1#2{0.9#1}
\epsfbox[100 580 500 700]{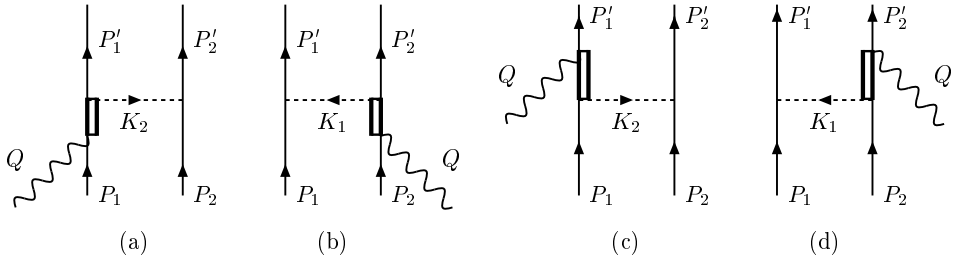}
\end{center} 
\caption{The Feynman diagrams corresponding to the two-body
electroexcitation of the $\Delta$.
}
\end{figure}

The structure of the work is the following: in sect.~2 we present the
relativistic model for the $\Delta$ current and provide expressions
for the ph matrix elements in the RFG. In sect.~3 we present the
results of the calculation of the nuclear response functions including
the $\Delta$ current for several values of the momentum transfer, up to $q=3$
GeV/c. 
We make contact with the
scaling properties of the response functions and compare the
results with those obtained in non-relativistic approaches for low and
intermediate momentum transfer.
Finally in sect.~4 we present our conclusions.  
In appendix A we
discuss the properties of the $\Delta$ propagator and provide a new
method to derive its general form 
fulfilling the point transformation.
In Appendix B we give details on the 
non-relativistic reduction of the $\Delta$ current.

\section{The relativistic $\Delta$-isobar current}

In line with refs.~\cite{Ama02a,Ama02b} we perform our analysis in
first-order perturbation theory, namely we consider the contributions
arising from diagrams with only one pionic line --- our conventions
are discussed at length in the references cited.  We then evaluate the
contribution of a virtual $\Delta$ to the longitudinal and transverse
response functions
\begin{eqnarray}
R^L(q,\omega)&=&\left(\frac{q^2}{Q^2}\right)^2\left[
W^{00}-\frac{\omega}{q}(W^{03}+W^{30})+\frac{\omega^2}{q^2}W^{33}
\right]\label{eq2a} \\
R^T(q,\omega)&=&W^{11}+W^{22} \, ,
\label{eq2b}
\end{eqnarray}
linked to the hadronic tensor $W^{\mu\nu}$. In the symmetric ($Z=N$)
RFG model the one-body-$\Delta$ interference contribution to the
hadronic tensor reads\footnote{We assume $q>2k_F$, where $k_F$ is the
Fermi momentum, so that no Pauli blocking is present.}
\begin{equation}
 W_{OB-\Delta}^{\mu\nu}=
\frac{3Z}{8\pi k_F^3q} 
\int_{h_0}^{k_F} h dh E_{\np}
\int_0^{2\pi}d\phi_h
\sum_{s_p,s_h} 2{\rm Re}\, \left[\frac{m_N^2}{E_{\np}E_{\nh}}
j^\mu_{OB}(\np,\nh)^* 
j^\nu_\Delta(\np,\nh)\right] \, ,
\label{eq38} 
\end{equation}
where $(E_{\np},\np)=P^\mu=(H+Q)^\mu=(E_{\nh}+\omega,\nh+\nq)$
and the lower limit of the integral, $h_0$, 
is the minimum momentum required for a nucleon to participate
in the process (see refs.~\cite{Ama02b,Alv01} 
for details and explicit expressions).
The ph one-body EM current is given by 
\begin{equation}
j^\mu_{OB}(\np,\nh) = \overline{u}({\bf p})
\left(F_1\gamma^\mu+i\frac{F_2}{2m_N}\sigma^{\mu\nu}Q_\nu\right)
u({\bf h}) \, ,
\end{equation} 
where
$F_1$ and $F_2$ are the Dirac and
Pauli form factors and $ u({\bf h})\equiv u({\bf h},s_h,t_h)$ the free
Dirac spinor. 
The current $j^\nu_\Delta(\np,\nh)$,
associated with the diagrams of fig.~2,
is linked to the ph matrix element of the $\Delta$ current operator,
$\hat{j}^\nu_\Delta(Q)$  through
\begin{eqnarray}
\langle ph^{-1}|\hat{j}^\nu_\Delta(Q)|F\rangle
&=&
(2\pi)^3\delta^3(\nq+\nh-\np)
\frac{m_N}{V\sqrt{E_{\np}E_{\nh}}}
j^\nu_\Delta(\np,\nh)
\label{phfunction}\\
&=&
\sum_{s_k,t_k}\sum_{\nk \leq k_F}
           \left[\langle pk|{\hat j}^\nu_\Delta|h k\rangle -
                 \langle pk|{\hat j}^\nu_\Delta|kh\rangle \right] \, .
\label{eq22}
\end{eqnarray}
The sum $\sum_{\nk \leq k_F}$ becomes,
in the thermodynamic limit,
an integral over the momentum in the range $0\leq k\leq k_F$, 
and over the angular variables
$\theta_k,\phi_k$.  The first and second terms in eq.~(\ref{eq22})
represent the direct and exchange contribution to the matrix element,
respectively.

\begin{figure}[t]
\begin{center}
\leavevmode
\def\epsfsize#1#2{0.9#1}
\epsfbox[100 590 500 700]{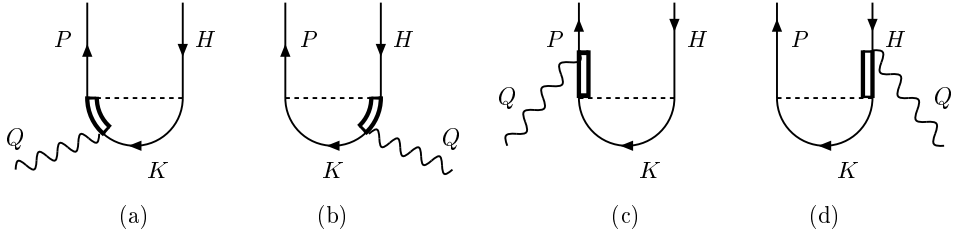}
\end{center}
\caption{The particle-hole Feynman diagrams corresponding to the two-body
electroexcitation of the $\Delta$.
}
\end{figure}

We may write the $\gamma N\Delta$ Lagrangian in the general form~\cite{Pas95}
\begin{equation}
{\cal L}_{\gamma N\Delta}= 
{\cal L}^1_{\gamma N\Delta}+
{\cal L}^2_{\gamma N\Delta}+ 
{\cal L}^3_{\gamma N\Delta} 
\end{equation}
with
\begin{eqnarray}
{\cal L}^1_{\gamma N\Delta} 
&=& \frac{ie G_1}{2 m_N}
\bar\psi^\alpha \Theta_{\alpha\mu}(z_1,A) \gamma_\nu 
\gamma_5 T_3^{\dagger} N
F^{\nu\mu} + \mbox{h.c.} 
\label{lag1}
\\
{\cal L}^2_{\gamma N\Delta} 
&=& \frac{eG_2}{(2 m_N)^2}
\bar\psi^\alpha \Theta_{\alpha\mu}(z_2,A) 
\gamma_5 T_3^{\dagger} (\partial_\nu N)
F^{\nu\mu} + \mbox{h.c.}
\label{lag2}
\\
{\cal L}^3_{\gamma N\Delta} 
&=& 
-\frac{eG_3}{(2 m_N)^2}
 \bar\psi^\alpha \Theta_{\alpha\mu}(z_3,A) \gamma_5 T_3^{\dagger} N
 \partial_\nu F^{\nu\mu} + \mbox{h.c.}\ ,
\label{lag3}
\end{eqnarray}
where $\psi^\alpha$ is the $\Delta$-field, $N$ the nucleon field and
$F^{\nu\mu}$ the EM field tensor.  We denote by
$T_a^{\dagger}$ the $\frac12 \rightarrow \frac32$ isospin transition
operators~\cite{Gil97} for $a=1,2,3$. The Lagrangians
(\ref{lag1}-\ref{lag3}) coincide with
the expressions given in refs.~\cite{Ben89,Dav91,Duf68}.  On the
contrary, the term in eq.~(\ref{lag2}) differs from ref.~\cite{Pas95} in a
global sign.  As it will be shown later, this difference makes a
negligible effect on the global contribution to the MEC($\Delta$)
transverse response, but it causes a very significant reduction in the
contribution to the longitudinal response.

The tensor $\Theta_{\mu\nu}(z,A)$ can be written in the general form
\be
\Theta_{\mu\nu}(z,A)=g_{\mu\nu}+\left[z+\frac{1}{2}(1+4z)A\right]
\gamma_\mu \gamma_\nu \, ,
\label{Theta}
\ee
where $z$ is the so-called off-shell parameter and $A$ is an arbitrary
parameter related to the ``contact'' invariance of the
Lagrangian~\cite{Ben89,Pas95}.

To derive the $\Delta$ current we need also the $\pi N \Delta$
Lagrangian, given by
\begin{equation}
{\cal L}_{\pi N\Delta}=
\frac{f_{\pi N\Delta}}{m_\pi}
\overline{\psi}^\mu\Theta_{\mu\nu}(z_\pi,A)\partial^\nu\npi\cdot\nT^{\dagger}N
\, ,
\end{equation} 
where $\npi$ is the isovector pion field, and 
we have introduced  an off-shell parameter $z_\pi$ for the 
$\pi N \Delta$ vertex. 
Finally, as in ref.~\cite{Ama02b}, 
we use pseudo-vector coupling for the $\pi NN$ vertex.

The corresponding two-body Delta current is obtained by computing the
$S$-matrix element for the elementary virtual photo-absorption process
by two nucleons $N_1+N_2+\gamma\rightarrow N'_1+N'_2$ 
(see fig.~1). The corresponding current function
can be written as\footnote{Here we use the Bjorken and Drell
conventions~\cite{Bjo65}, whereas different conventions were used in
\cite{Van81}.} (see ref.~\cite{Ama02b} for a detailed definition of
relativistic currents):
\begin{eqnarray}
j^\mu_\Delta(\np'_1,\np'_2,\np_1,\np_2)
&=& 
\frac{f_{\pi N\Delta} f}{m_\pi^2} 
G_\pi(K_2) K_{2\alpha} Q_\nu 
\overline{u}({\bf p'_1})
\left[ X_a^{\alpha\mu\nu}(P'_1,P_1)-X_a^{\alpha\nu\mu}(P'_1,P_1)\right] 
u({\bf p_1})
\nonumber\\
&&\times
\overline{u}({\bf p'_2}) \gamma_5 \Kbar_2 
\tau_a u({\bf p_2}) 
+ (1\longleftrightarrow 2)\, ,
\label{jmuD}
\end{eqnarray}
where we use the Einstein convention for the Lorentz indices and for
a sum over a repeated isospin index $a=1,2,3$.
Moreover, $K^\mu_i=P'_i{}^\mu-P^\mu_i$ (with $i=1,2$)
are the pionic four-momenta (see fig.~1) and
\begin{equation}
G_\pi(K) = \frac{1}{K^2-m_\pi^2}
\end{equation}
is the propagator of a pion carrying four-momentum $K^\mu$.
The  tensor  $X_a^{\alpha\mu\nu}$ is defined as

\begin{eqnarray}
\lefteqn{X_a^{\alpha\mu\nu}(P',P) =
\Theta^{\alpha\beta}(z_\pi,A) G^\Delta_{\beta\rho}(P+Q)}
\nonumber\\
&\times&
\left[
       \frac{G_1}{2 m_N}   \Theta^{\rho\mu}(z_1,A)\gamma^\nu
     - \frac{G_2}{4 m_N^2} \Theta^{\rho\mu}(z_2,A)P^\nu
      +\frac{G_3}{4 m_N^2} \Theta^{\rho\mu}(z_3,A)Q^\nu
\right]  
\gamma_5 T_a T_3^{\dagger} 
\nonumber\\
&+&
\gamma_5 
\left[
       \frac{G_1}{2 m_N} \gamma^\nu \Theta^{\mu\rho}(z_1,A)
      -\frac{G_2}{4 m_N^2} P^{'\nu} \Theta^{\mu\rho}(z_2,A)
      -\frac{G_3}{4 m_N^2}Q^\nu     \Theta^{\mu\rho}(z_3,A)
\right] 
\nonumber\\
&\times&
G^\Delta_{\rho\beta}(P'-Q) \Theta^{\beta\alpha}(z_\pi,A)
T_3 T_a^{\dagger}\, .
\label{jmuD1}
\end{eqnarray}
The isospin sums in eq.~(\ref{jmuD}) can be performed using the relations
\begin{eqnarray}
T_a^{(1)} T_3^{\dagger(1)} \tau^{(2)}_a 
&=& 
\frac{2}{3} \tau_z^{(2)}
- \frac{i}{3} \left[\tauvec^{(1)} \times \tauvec ^{(2)} \right]_z
\label{iso}
\\
T_3^{(1)} T_a^{\dagger(1)} \tau^{(2)}_a 
&=& 
\frac{2}{3} \tau_z^{(2)}
+\frac{i}{3} \left[\tauvec^{(1)} \times \tauvec ^{(2)} \right]_z \ .
\label{iso2}
\end{eqnarray}
In eq.~(\ref{jmuD1}) the three amplitudes contributing to the
N-$\Delta$ vertex~\cite{Jon73,Pas95} are taken into account and the
values of the off-shell parameters $z_\pi$, $z_1$, $z_2$ and $z_3$
will be discussed later.

In what follows we start by considering the $\Delta$ as a stable
particle with mass $m_\Delta$, and we will later include a width to
account for its decay probability in the resonance region.  The isobar
propagator can be expressed in general as a sum of two terms 
\begin{equation}
G^\Delta_{\beta\rho}(P) =
G^{RS}_{\beta\rho}(P) + G^A_{\beta\rho}(P) \, ,
\end{equation}
where
\begin{equation}
 G^{RS}_{\beta\rho}(P) 
= -\frac{ \Pbar+m_\Delta}{P^2-m_\Delta^2}
\left[  
         g_{\beta\rho} 
        - \frac{1}{3} \gamma_\beta\gamma_\rho 
        - \frac{2}{3} \frac{P_\beta P_\rho}{m_\Delta^2} 
        - \frac{\gamma_\beta P_\rho - \gamma_\rho P_\beta}{3m_\Delta} 
\right]
\label{Dprop}
\end{equation}
is the usual Rarita-Schwinger (RS) propagator tensor and 
\begin{eqnarray}
G^A_{\beta\rho}(P) 
&=&
-\frac{1}{3m_\Delta^2}\frac{A+1}{(2A+1)^2} 
\nonumber \\
&&
\mbox{}\times 
\left[
        (2A+1)(\gamma_\beta P_\rho + P_\beta\gamma_\rho)
       -\frac{A+1}{2}\gamma_\beta (\Pbar +2m_\Delta)\gamma_\rho 
       + m_\Delta\gamma_\beta\gamma_\rho
\right] 
\label{Aprop}
\end{eqnarray}
the piece of the propagator that depends on the parameter $A$.
Note that the global sign in the RS term differs from the expressions
given in \cite{Ben89,Dav91,Pas95}.  This can be due just to a
different choice of phase in the definition of the propagator.  As
shown in Appendix A, the choice of phase in this work coincides
with the one that provides
the standard form of the nucleon propagator, $S_N(P)=1/(\Pbar-m_N)$.
In appendix A 
eq.~(\ref{Aprop}) is derived using a contact transformation of 
the RS propagator\footnote{
Note that there is an error in the relative sign between
the $G^{RS}_{\beta\rho}$ and  $G^A_{\beta\rho}$ pieces given in 
refs.~\cite{Ben89,Hab98}.}.
Moreover, because of the contact
invariance, the physical properties of the field can be shown not to
depend on $A$. 
As a further test of our calculation, we have checked that the results
for the $T$ response do not depend on $A$.
Hence in what follows we fix $A=-1$ so that the
complete $\Delta$ propagator is simply reduced to the Rarita-Schwinger
expression and therefore omit the explicit $A$-dependence in the tensor
$\Theta^{\mu\nu}$.

It is immediate to check that the current in eq.~(\ref{jmuD}) is conserved,
\begin{equation}
Q_\mu j^\mu_\Delta(\np'_1,\np'_2,\np_1,\np_2)=0 \, ,
\end{equation}
so that the $\Delta$ current 
does not affect the gauge invariance of the RFG model.

The particle-hole matrix-element diagrams are obtained from eq.~(\ref{jmuD})
by setting 
\ba
(P'_1,s'_1,t'_1) &=& (P,s_p,t_p)\ ,\ \ \ 
(P_1,s_1,t_1) = (H,s_h,t_h)
\nonumber\\
(P_2,s_2,t_2) &=& (P'_2,s'_2,t'_2) = (K,s_k,t_k)
\ea
for the {\em direct} term and
\ba
(P'_1,s'_1,t'_1) &=& (P,s_p,t_p)\ ,\ \ \ 
(P_2,s_2,t_2) = (H,s_h,t_h)
\nonumber\\
(P_1,s_1,t_1) &=& (P'_2,s'_2,t'_2) = (K,s_k,t_k)
\ea
for the {\em exchange} term, and by summing over the hole momentum $k$,
spin $s_k$ and isospin $t_k$.
The isospin trace yields a vanishing direct matrix element, since 
${\rm Tr}\,\tau_a = {\rm Tr}\,T_3^{\dagger} T_a = 0$.
In the exchange channel, recalling eq.~(\ref{iso}), we get
\begin{equation}
\sum_{t_k} \chi^{\dagger}_{t_p} T_a T_3^{\dagger}
\chi_{t_k} \chi^{\dagger}_{t_k} \tau_a \chi_{t_h}
=\sum_{t_k} \chi^{\dagger}_{t_k} T_3 T_a^{\dagger}
 \chi_{t_h} \chi^{\dagger}_{t_p} \tau_a \chi_{t_k}
= \frac{4}{3} \chi^{\dagger}_{t_p} \tau_z \chi_{t_h} 
\end{equation}
and
\begin{equation}
\sum_{t_k} \chi^{\dagger}_{t_p} T_3 T_a^{\dagger} 
\chi_{t_k} \chi^{\dagger}_{t_k} \tau_a \chi_{t_h}
=\sum_{t_k} \chi^{\dagger}_{t_k} T_a T_3^{\dagger} 
\chi_{t_h} \chi^{\dagger}_{t_p} \tau_a \chi_{t_h}=
0\ .
\end{equation}
As a consequence only the two ``vertex'' diagrams (a) and (b) in
fig.~2, contribute to the process, whereas the ``self-energy''
diagrams (c) and (d) give a vanishing contribution to the responses.

Finally, by taking the thermodynamic limit, we get for the 
ph current function in eq.~(\ref{phfunction}):
\begin{eqnarray}
j^\mu_\Delta(\np,\nh)
&=&
  -\frac{f_{\pi N\Delta} f}{m_\pi^2} 
   \frac{4}{3} \chi^{\dag}_{t_p} \tau_z \chi_{t_h} 
   \sum_{s_k} \int\frac{d^3 k}{(2\pi)^3} \frac{m_N}{E_{\nk}}\theta(k_F-k) 
\nonumber\\
&&
\mbox{}\times
\left[\right. 
\overline{u}({\bf p},s_p) F^\mu(P,H,K) u({\bf k},s_k) 
\overline{u}({\bf k},s_k) \gamma_5 (\Kbar-\Hbar) u({\bf h},s_h)
\nonumber\\
&& 
\kern 1em\mbox{}
+\overline{u}({\bf p},s_p) \gamma_5 (\Pbar-\Kbar) u({\bf k},s_k)
\overline{u}({\bf k},s_k) B^\mu(P,H,K) u({\bf h},s_h)
\left.\right] \, ,
\label{phme} 
\end{eqnarray}
where 
\begin{eqnarray}
F^\mu(P,H,K) 
&\equiv& 
   G_\pi(K-H) (K-H)_\alpha Q_\nu
   \Theta^{\alpha\beta}(z_\pi) G^\Delta_{\beta\rho}(K+Q)
\nonumber \\ 
&&
\mbox{}\times
\left\{\frac{G_1}{2 m_N} 
       \left[\Theta^{\rho\mu}(z_1)\gamma^\nu-\Theta^{\rho\nu}(z_1)\gamma^\mu
       \right]
      -\frac{G_2}{4 m_N^2} 
      \left[\Theta^{\rho\mu}(z_2) K^\nu-\Theta^{\rho\nu}(z_2) K^\mu
      \right]
\right.
\nonumber\\
&& \left.
   \mbox{} +\frac{G_3}{4 m_N^2} 
       \left[\Theta^{\rho\mu}(z_3) Q^\nu-\Theta^{\rho\nu}(z_3) Q^\mu
      \right]
\right\}\gamma_5
\end{eqnarray}
and
\begin{eqnarray}
\lefteqn{
  B^\mu(P,H,K) \equiv
  G_\pi(P-K) (P-K)_\alpha Q_\nu 
}
\nonumber\\
&&
\mbox{}\times
\gamma_5 
\left\{ \frac{G_1}{2 m_N} 
        \left[\gamma^\nu \Theta^{\mu\rho}(z_1)-\gamma^\mu \Theta^{\nu\rho}(z_1)
        \right]
       -\frac{G_2}{4 m_N^2}
        \left[ K^\nu \Theta^{\mu\rho}(z_2)- K^\mu\Theta^{\nu\rho}(z_2)
        \right] 
\right.
\nonumber\\
&&
\left.\mbox{}
      -\frac{G_3}{4 m_N^2}
       \left[Q^\nu \Theta^{\mu\rho}(z_3) -Q^\mu\Theta^{\nu\rho}(z_3) 
       \right]
\right\}
G^\Delta_{\rho\beta}(K-Q) \Theta^{\beta\alpha}(z_\pi)
\end{eqnarray}
correspond to the forward- and backward-going diagrams (a) and (b) of
fig.~2, respectively.

The 
decay of the $\Delta$-resonance into a physical $N$-$\pi$
state should be taken into account above threshold, i.e.,
$P_\Delta^2>(m_\pi+m_N)^2$.  This is accounted for by modifying the
propagator in eq.~(\ref{Dprop}) to include a finite width according to the
following prescription~\cite{Dekker}
\begin{equation}
\frac{1}{P^2-m_\Delta^2} \longrightarrow
\frac{1}{P^2-\left[m_\Delta-\frac{i}{2}\Gamma(P^2)\right]^2} 
\end{equation}
in which~\cite{Gil97}
\begin{equation}
\Gamma(P^2) = \Gamma_0 
            \frac{m_\Delta}{\sqrt{P^2}}
            \left(\frac{p_\pi^*}{p_\pi^{res}}\right)^3 
\end{equation}
is the energy-dependent width. In the above 
$p_\pi^*$ is the momentum of the final pion resulting from the 
$\Delta$ decay (in the $\Delta$-system)
and $p_\pi^{res}$ is its value at resonance.
Moreover we take $\Gamma_0=120$ MeV.


\section{Results}


In this section we show the contribution of the $\Delta$  current
to the longitudinal and transverse quasielastic response functions
in the 1p-1h channel.

\subsection{Response functions and off-shell dependence}

The following coupling constants have been used in the 
calculation~\cite{Dekker,Pas95}:
\be
f=\sqrt{4\pi\times0.08}\ ,\  
f_{\pi N\Delta}=4\times 0.564\ ,\      
G_1=4.2\ ,\                      
G_2=4\ ,\                        
G_3=1  \ .                      
\ee
The value of $G_3$ is arbitrary owing to the lack of experimental 
information (see below).
Moreover, although not explicitly indicated in the above formulae,
the following monopole form factor
\begin{equation}
F_{\pi NN}(K) = F_{\pi N\Delta}(K)=
\frac{\Lambda^2-m_\pi^2}{\Lambda^2-K^2}\ ,
\end{equation}
with $\Lambda=1300$ MeV, has been used. For the single-nucleon current
we have for simplicity adopted the Galster form factor 
parameterization~\cite{Gal71}. The Fermi momentum is chosen to be
$k_F=237$ MeV/c, namely, representative of a typical sd-shell
nucleus. 

\begin{figure}[tp]
\begin{center}
\leavevmode
\def\epsfsize#1#2{0.9#1}
\epsfbox[100 265 500 745]{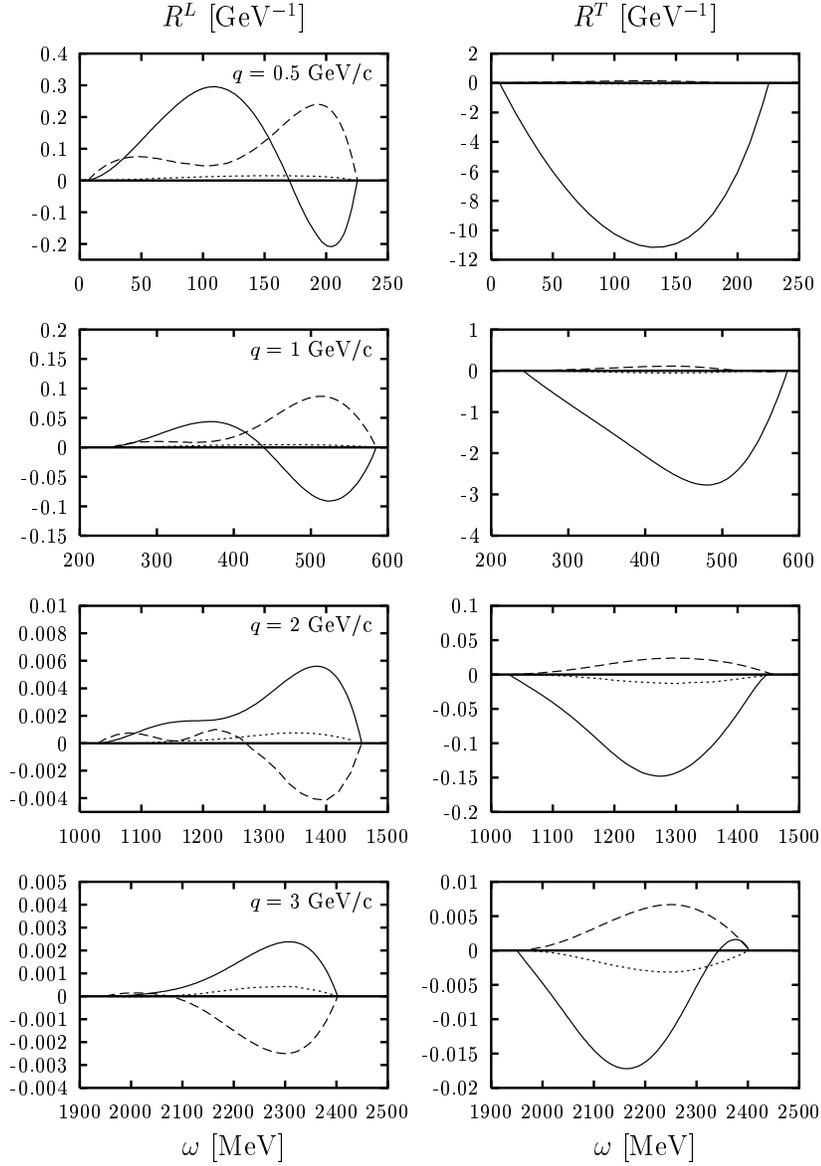}
\end{center}
\caption{The contribution of the $\Delta$ current to the
longitudinal (left panels) and transverse (right panels) responses 
plotted versus $\omega$. Here $k_F$=237 MeV/c.
The separate contributions of the first ($G_1$, solid) 
second ($G_2$, dashed) and third ($G_3$, dotted) terms 
of the current are displayed.
The off-shell parameters are taken as $z_1=z_2=z_3=z_\pi=-1/4$.
}
\end{figure}

In fig.~3 we plot the one-body-$\Delta$ interference contributions to
the longitudinal and transverse responses as functions of the energy
transfer $\omega$ for various values of the momentum transfer $q$,
ranging from 0.5 to 3 GeV/c.  The separate contributions of the
three Lagrangians in eqs.~(\ref{lag1},\ref{lag2},\ref{lag3}) are displayed.

Regarding the longitudinal response (left panels), the $G_1$ and $G_2$
pieces are similar in magnitude and tend to cancel for high $q$,
whereas the $G_3$ term is negligible. Note that this result is very
different from the one obtained with the Peccei Lagrangian
(corresponding to the $G_1$ term only). However, as will be shown
later, the whole contribution to the response is very small.

In the transverse channel (right panels) the $G_1$ term clearly
dominates, although at high $q$ the contribution of $G_2$ becomes
significant and tends to cancel the first one. Hence in this case the
results almost coincide, for $q$ not too high, with the results
obtained with the Peccei Lagrangian.

\begin{figure}[tp]
\begin{center}
\leavevmode 
\def\epsfsize#1#2{0.9#1}
\epsfbox[100 500 500 740]{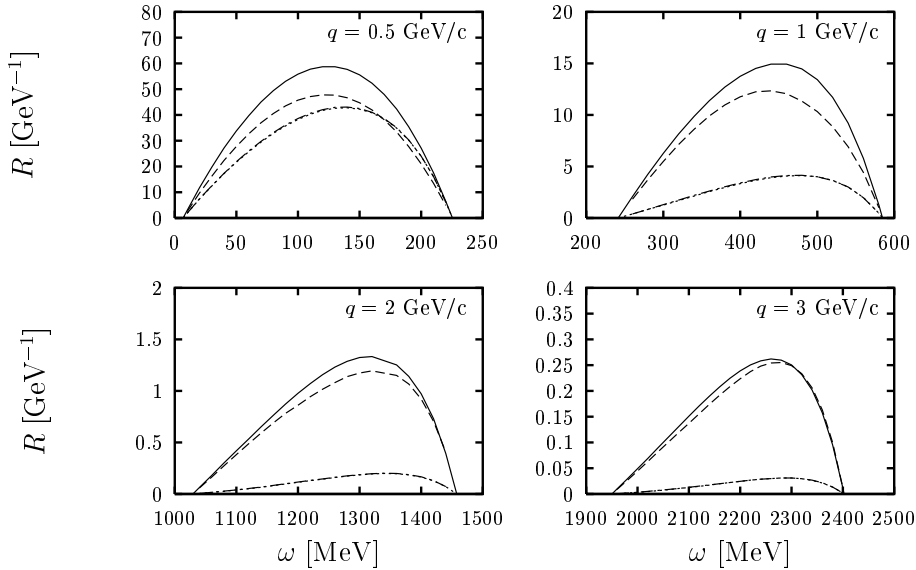}
\end{center} 
\caption{Global longitudinal and transverse responses plotted versus
$\omega$. Solid: RFG transverse response;
dashed: RFG+MEC($\Delta$) transverse response;
dotted: RFG longitudinal response;
dot-dashed: RFG+MEC($\Delta$) longitudinal response.
}
\end{figure}

\begin{figure}[tp]
\begin{center}
\leavevmode 
\def\epsfsize#1#2{0.9#1}
\epsfbox[100 490 500 730]{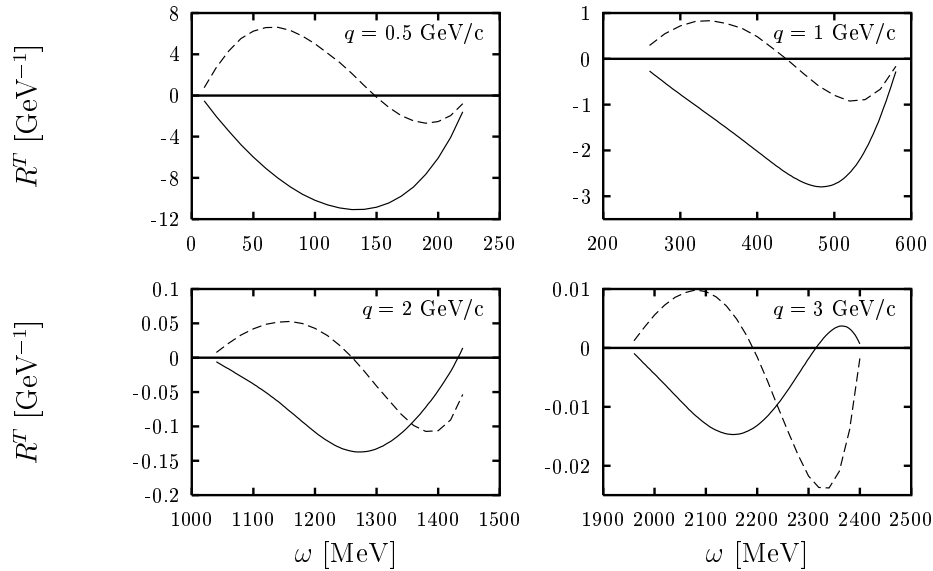}
\end{center} 
\caption{MEC($\Delta$) contribution to the transverse response (solid)
compared with the pionic seagull plus pion-in-flight contribution
(dashed).  }
\end{figure}

To clarify the effects introduced by the $\Delta$ current, we show in
fig.~4 the ``total'', namely RFG+MEC($\Delta$), longitudinal and
transverse responses compared with the pure RFG ones.  The same values
of $q$ as in the previous figure have been considered.  We note that
the contribution of the MEC($\Delta$) to the longitudinal response is
always negligible.  In the transverse channel the contribution of the
$\Delta$ is larger and negative, that is, the interference between the
one- and two-body currents matrix elements is destructive and
therefore reduces the total from the purely nucleonic answer.  At low
momentum transfers the fractional contribution arising from
MEC($\Delta$) contributions is relatively large and then it slowly
decreases as the momentum transfer increases into the several GeV
regime. Specifically, at the selected values of $q=$ 0.5, 1, 2 and 3
GeV/c the net effect of the MEC($\Delta$) contributions is 19\%, 18\%,
10\% and 4\%, respectively.

It is also interesting to compare the role of the MEC($\Delta$) with
the other pionic MEC (seagull and pion-in-flight), calculated in the
same relativistic model in \cite{Ama02a,Ama02b}. As shown in fig.~5
the $\Delta$ contribution is larger than the pionic one for lower $q$,
where the total MEC are sizable, although for higher momentum
transfers the difference in magnitude between $\Delta$ and pionic MEC
contributions decreases and the two contributions tend to cancel. As a
consequence, the impact of the total MEC at high $q$ (say $q\geq 2$
GeV/c) is almost vanishing in the 1p-1h sector.

\begin{figure}[tp]
\begin{center}
\leavevmode 
\def\epsfsize#1#2{0.9#1}
\epsfbox[50 340 480 730]{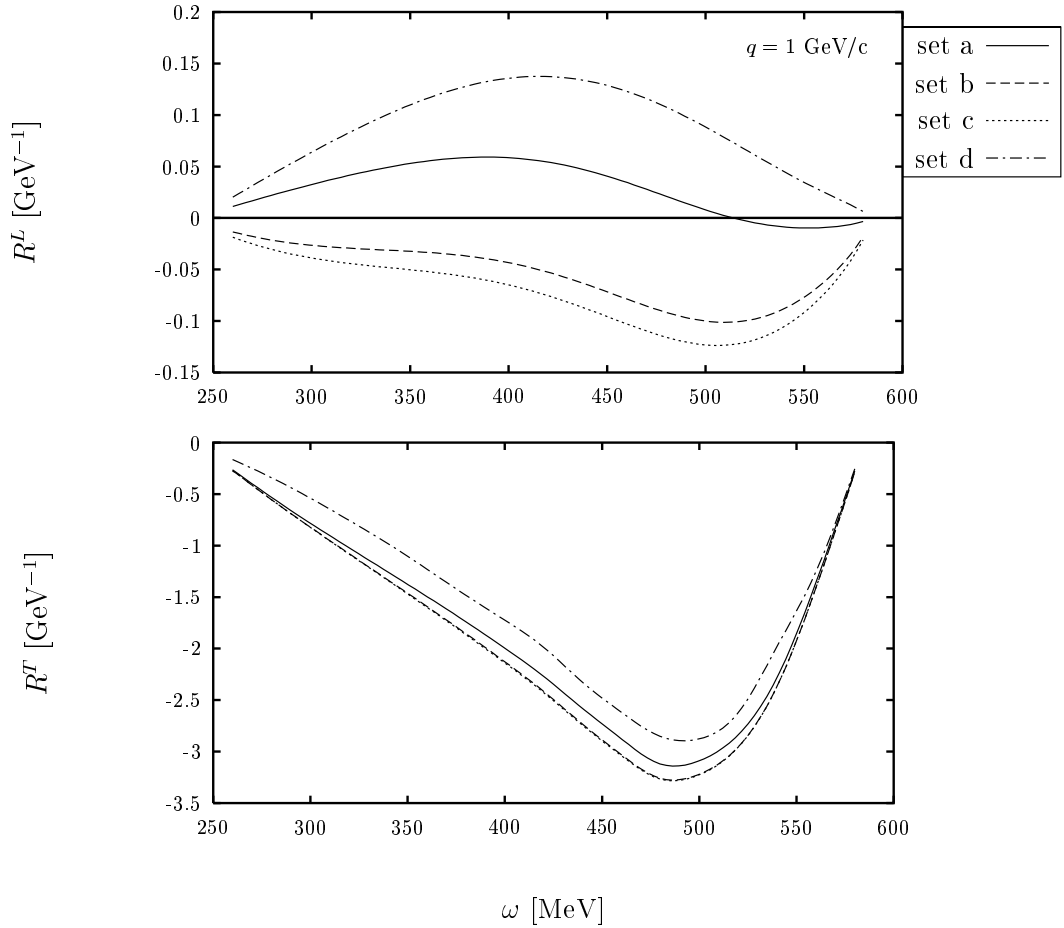}
\end{center}
\caption{MEC($\Delta$) contribution to the longitudinal (upper panel)
and and transverse (lower panel) responses plotted versus
$\omega$. The momentum transfer is $q=1$ GeV/c and results are shown
for the four sets of the off-shell parameters discussed in the text.}
\end{figure}

We now investigate the effect of changing the off-shell parameters
$z_i$.  The values
$z_i=-0.25$ used in the above results are unnecessarily restrictive, since
they were fixed by Peccei under the condition
$\gamma_\mu\Theta^{\mu\nu}(z,A)=0$, which does not correspond to the
most general form
of $\Theta_{\mu\nu}$ consistent with the point transformation.
Different ranges or sets of the parameters $z_\pi$, $z_1$ and $z_2$ have been 
obtained by fitting pion photo-production data on the nucleon~\cite{Dav91,Dav86},
while the value of $z_3$  needs electrons to be fixed.
The uncertainty in these parameters arises from
the treatment of the unitarity constraint,
due to both theoretical ambiguities and to lack of precise experimental data.

Since a unique determination of the off-shell
parameters is not available, we have studied the dependence of our results 
upon a variation of them. 
In fig.~6 we present the $\Delta$ contribution to the
longitudinal and transverse responses for $q=1$ GeV/c 
using four different choices of
the parameters $(z_\pi,z_1,z_2,z_3)$:
\begin{eqnarray}
& a)\  \left(-1/4,-1/4,-1/4,-1/4\right)\ ,
& b)\  \left(-1/4,0.1,2.25,-1/4\right)\ , 
\nonumber\\
& c)\  \left(-1/4,0.3,2.25,-1/4\right)\ ,
& d)\  \left(-1/2,-1/2,-1/2,-1/2\right)\ .
\label{off}
\end{eqnarray}
Set a), used in the previous figures, reduces to the standard Peccei
Lagrangian in the case $G_2=G_3=0$.  Sets b) and c) are determined
in \cite{Pas95} by fitting Compton scattering cross sections on the
nucleon (note that in this case the $G_3$ term does not contribute)
and set d) yields the usual non-relativistic $\gamma N\Delta$ vertex,
namely $\Theta_{\mu\nu}=g_{\mu\nu}$.  
Although several other choices of
parameters have been suggested in the literature~\cite{Dav91}, 
we believe that the results of fig.~6 give an
indication of the uncertainty in the response functions associated with
different off-shell prescriptions.  It appears
from fig.~6 that the fractional uncertainty is much larger in the longitudinal
channel, where, however, the contribution to the response is
negligible.  On the other hand the transverse response, for which the
MEC($\Delta$) contributions are more important, is rather
insensitive to the different 
choices (the effect being at most of the order of 6--7\%).

   In~\cite{Pas01} a correspondence between classes of ``consistent'' and
   ``inconsistent'' lagrangians was found by a redefinition of the
   spin-3/2 field. A new contact interaction that does not involve
   the spin-3/2 field appears. In our case this would mean that a new
   MEC term of contact type, dependent on the off-shell parameters,
   would arise and should be added to the "seagull" current,
   cancelling  the dependence of the total responses on the off-shell
   parameters.
   
In the next subsection we shortly address the scaling behavior of the
MEC($\Delta$) responses.

\subsection{Scaling}

The scaling phenomenon has been presented in detail in
refs.~\cite{Don99a,Don99b,Mai02}. Here we just recall the basic
definitions which are of use for the discussion that follows.  We only
focus on the transverse channel since, as previously shown, the
MEC($\Delta$) are irrelevant in the longitudinal channel.

Scaling of first kind occurs if the scaling function
\be
f^T(q,\omega)=k_F \frac{R^T(q,\omega)}{G^T(q,\omega)}
\label{fLT}
\ee becomes a function of one single variable,
the scaling variable, and independent of $q$.  Such behaviour is
known to occur for large $q$ in the region below the QEP.
In eq.~(\ref{fLT}) $G^T(q,\omega)$ is the
relevant single-nucleon EM function (see
ref.~\cite{Mai02} for its explicit expression).

Several different scaling variables exist in the literature, all of them
coalescing into one --- or being simply related to each other --- for high
enough momentum transfers. In the quasielastic peak region the 
natural scaling variable turns out to be~\cite{Alb93,Bar98}
\be
\psi=\pm\sqrt{\frac{T_0}{T_F}}\ ,
\label{psi}
\ee
where $T_0=\sqrt{h_0^2+m_N^2}-m_N$
is the minimum kinetic energy required to a nucleon to take part in the
process. The $+(-)$ sign in 
eq.~(\ref{psi}) refers to the right (left) of the quasielastic peak.
The analysis of the World data~\cite{Don99a,Don99b,Mai02} 
shows that scaling of first kind is reasonably
good for $\psi<0$ and badly violated for $\psi>0$.

Scaling of second kind corresponds to the independence of the function
$f^T$ on the specific nucleus, namely on the Fermi 
momentum. The analysis of the existing data points to an excellent 
fulfillment of this scaling in the region $\psi<0$ and to a not very 
dramatic breaking of it for $\psi>0$.
When the two kinds of scaling occur the response is said to ``superscale''.

The relativistic Fermi gas model fulfills both kinds of scaling, by 
construction, yielding the scaling function $f_0=3(1-\psi^2)/4$.
The observed superscaling behavior of the experimental 
data~\cite{Don99a,Mai02} offers a clear 
constraint on the size allowed for nuclear correlations and MEC
contributions, since these may break the scaling behaviour (of both
kinds), and can therefore be used as a test of the 
reliability of the model.
It is then natural to explore the scaling behaviour predicted in 
the present model.

\begin{figure}[t]
\begin{center}
\leavevmode 
\def\epsfsize#1#2{0.9#1}
\epsfbox[50 410 480 730]{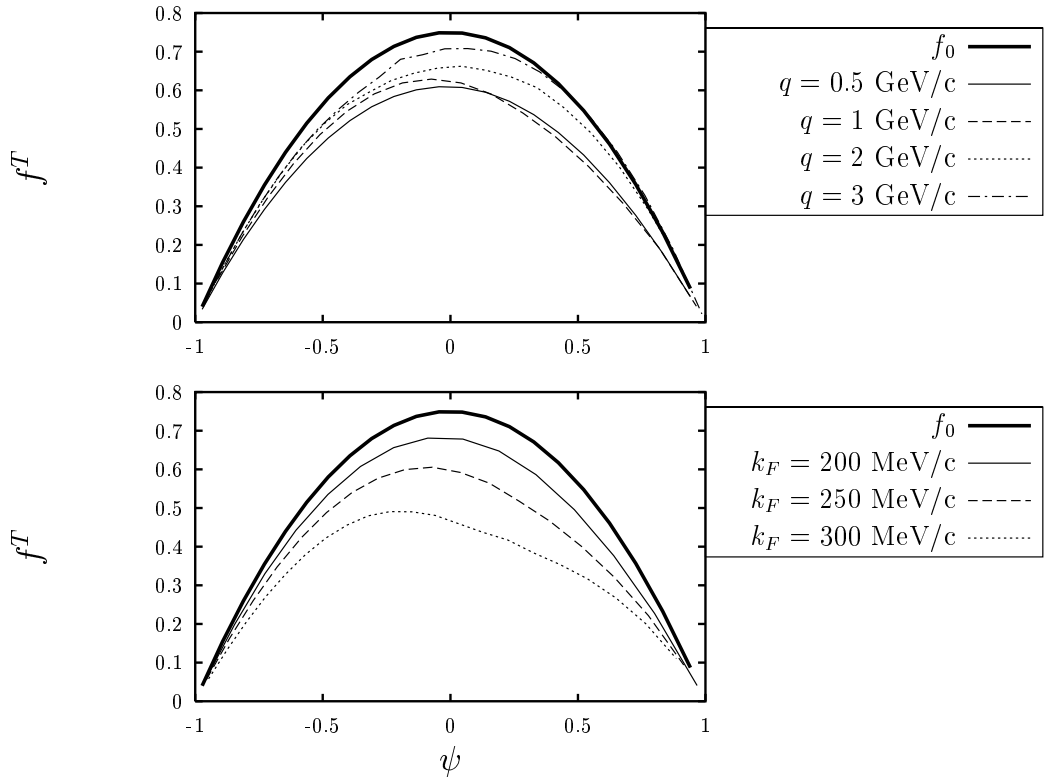}
\end{center}
\caption{The scaling function $f^T$ plotted versus $\psi$ at $k_F=237$
MeV/c for various values of $q$ (upper panel) and at $q=1$ GeV/c for
various values of $k_F$ (lower panel). The function $f_0$ (thick solid
line) refers to the free Fermi gas scaling function.  }
\end{figure}

In \cite{Ama02a} the evolution with $q$ and $k_F$ of
the pionic MEC and correlations has been already explored in detail:
it has been proven that these contributions 
satisfy, at momentum transfers above
1 GeV/c, scaling of the first kind, but that they violate
the second-kind scaling by roughly three powers of $k_F$, which is
strong enough to be seen in existing data. However, the size of the
scale breaking predicted in \cite{Ama02a} in the scaling region
(below the QEP) was small enough 
to be compatible 
with the high quality high-$q$ data.

In fig.~7 we display the scaling function $f^T$ versus $\psi$ for
different values of $q$ (upper panel) and $k_F$ (lower panel).  It
appears that both kinds of scaling are violated by the MEC($\Delta$)
contribution, something not evident in the high quality World data, at
least within existing experimental uncertainties.  When the ``total''
1p-1h response is formed a quantitative analysis shows that the
MEC($\Delta$) contributions in the transverse response play some role,
but that overall first-kind scaling is quite good for $-1\leq \psi
\leq -0.5$ and $q\leq 1-2$ GeV/c and is violated at a 5-10\% level at
the QEP. On the other hand, the second-kind scaling violations from
this 1p-1h model can be significant for all values of $\psi$, since
they modify the scaling function (that scales, by construction) by
three powers of $k_F$. The figure shows a generous range of Fermi
momenta; these correspond to going from $^4$He at $k_F=$ 200 MeV/c to
very heavy nuclei at $k_F\cong$ 250 MeV/c and then well beyond to 300
MeV/c to explore more fully the second-kind scaling behaviour. If we
restrict our attention to nuclear species where high quality data
exist then we should compare the $k_F=$ 200 MeV/c results with those
for $k_F=$ 250 MeV/c and the lower panel in the figure shows that
violations of about 14\% are predicted at the QEP (actually using
modeling of this type for helium is probably stretching things
somewhat: if the comparison is made for ``real nuclei'' such as
$^{12}$C versus $^{197}$Au for which excellent data exist then an
effect of roughly 8\% is predicted). While second-kind scale breaking
of this magnitude should be visible in the data one should be careful
before drawing premature conclusions, since one knows that additional
scale-breaking contributions arise from mechanisms outside the present
model. In particular we know from past work~\cite{Dekker,Van81,Alb84}
that 2p-2h MEC contributions also typically lead to effects that go as
$k_F^3$, but these {\em add} and therefore tend to cancel the
contributions discussed in the present work. The net effect, in the
scaling region at least, is expected to amount to a few percent and
therefore be consistent with the relevant World data.

It appears that the scaling behaviour of the MEC($\Delta$)
contributions is different from that of the single-nucleon RFG
response --- namely, if the scaling functions and variables are
defined as in past work to make the latter scale, then the two-body
MEC contributions in general do not.  First-kind scaling violations
(see the upper panel in fig. 7) arise because the one- and two-body
currents do not in general have the same $(q,\omega)$ dependences, and
only one choice can be made when defining the scaling
function. Namely, in standard treatments it is the one-body
(single-nucleon) cross section that is divided out to produce scaling
for the dominant impulse approximation contributions.  Similarly,
second-kind scaling violations arise when the density dependences of
the various contributions are different.

The high-quality World data present a mixed picture: for negative
values of $\psi$ both kinds of scaling are reasonably respected,
although scaling of the second kind appears to be better than scaling
of the first kind in this region.  For $\psi$ positive, scaling of the
first kind becomes quite bad, partially because pion production
including via the delta and other baryon resonances becomes important
and, for the reasons stated above, has a different $(q,\omega)$
dependence.  Additionally, the MEC($\Delta$) effects under study in
the present work can contribute to this first-kind scale breaking.
Were the latter to be the only contributions to be present along with
the one-body responses, then we would conclude from this study that
overall first-kind scaling is quite good for $-1\leq \psi \leq -0.5$
and $q\leq 1$--2 GeV/c, but could be violated by as much as 5--10\% at
the QEP.  However, the full assessment of which contributions produce
the observed result cannot be made before all effects are included,
namely from impulse approximation nucleonic currents, 1p-1h MEC
effects from pion-exchange contributions including those involving the
$\Delta$ (this work), 2p-2h MEC effects (work in progress) and meson
production.

Similar statements can be made with respect to second-kind scaling
violations.  The MEC($\Delta$) contributions studied in the present
work (see lower panel in fig. 7) can be significant for all values of
$\psi$, since they modify the scaling function (that scales, by
construction) by three powers of $k_F$.

\subsection{Non-relativistic reduction}

\begin{figure}[tp]
\begin{center}
\leavevmode 
\def\epsfsize#1#2{0.9#1}
\epsfbox[100 155 500 725]{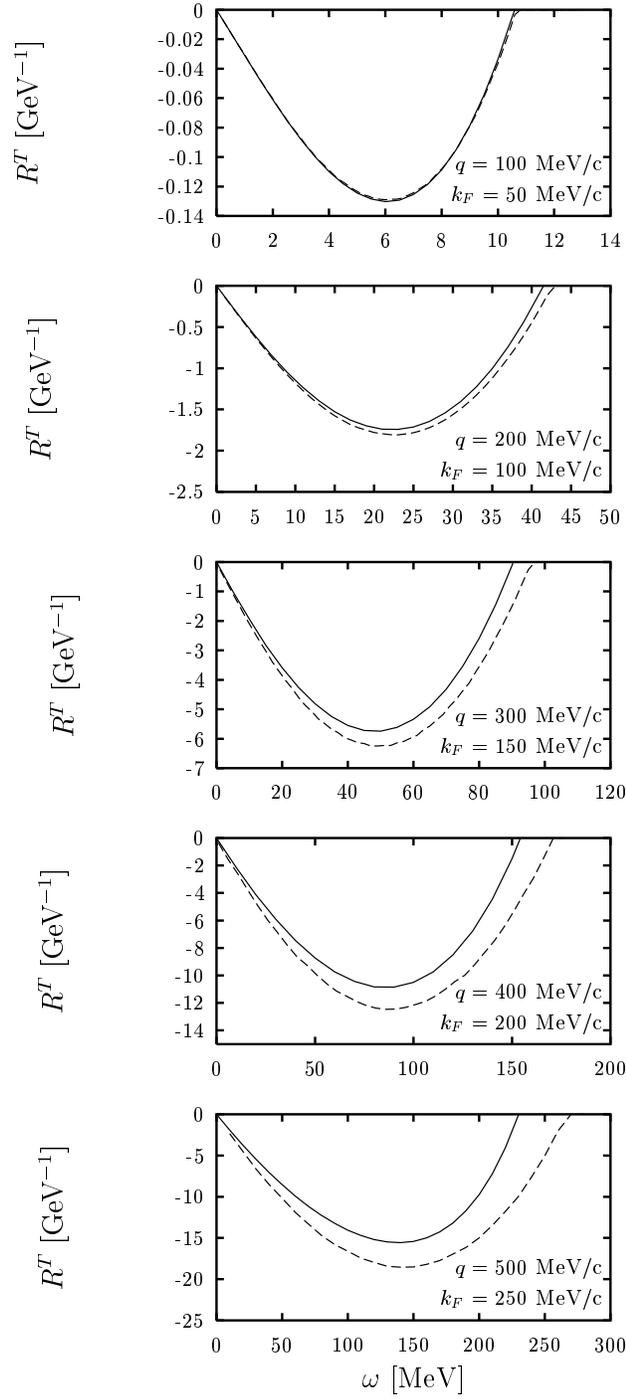}
\end{center}
\caption{MEC($\Delta$) transverse response computed for several values
of $q$ and $k_F$.  Solid: exact relativistic result, dashed:
non-relativistic result. Static propagators without a $\pi NN$ form
factor have been used.  }
\end{figure}

To finish with the discussion of the results, in this section we
analyze the effects associated with different types of non-relativistic
approaches.  Focusing on the MEC($\Delta$) contribution to the hadronic
$(e,e')$ observables one deals with the single-nucleon EM
current operator and the two-body $\Delta$ current. Improved
``semi-relativistic'' (SR) versions of the former have been presented
in \cite{Ama96a,Ama96b,Jes98} by expanding only in the dimensionless
parameter $\eta=p/m_N$, being $p$ the struck nucleon three-momentum.
By this method a SR form of the transverse OB current can be written 
in terms of the non-relativistic one as
\begin{equation}
\nj^{OB}_{SR}(\np,\nh)=\frac{1}{\sqrt{1+\tau}}\nj^{OB}_{NR}(\np,\nh)\, ,
\label{srob}
\end{equation}
where $\tau=|Q^2|/4m_N^2$.
A similar expansion has been carried out for the pion-in-flight and
contact  MEC in
ref.~\cite{Ama98} and for the one-body $\Delta$ electroproduction
current in \cite{Ama99}. 
In the case of the MEC, a semi-relativistic expression
similar to eq.~(\ref{srob}) was proposed and tested in \cite{Ama02b}
\begin{equation}
\nj^{MEC}_{SR}(\np,\nh)=\frac{1}{\sqrt{1+\tau}}\nj^{MEC}_{NR}(\np,\nh)\, .
\label{srmec}
\end{equation}
Although these SR MEC currents compare better than the NR ones
to the exact relativistic calculation, they are not as 
good as that the SR OB current since, as illustrated  
in \cite{Ama02b},
an additional $k_F$-dependent normalization factor $N(q,\omega,k_F)$,
arising from the integration over the Fermi
sea, should be present in eq.~(\ref{srmec}).
Nevertheless the simplicity of these SR currents make them suitable
for easy incorporation in existing non-relativistic
models.

The relativizing factor $(1+\tau)^{-1/2}$ takes care of
spinology and normalization properties that cannot be neglected 
in any relativistic model. Therefore, consistently
with eq.~(\ref{srmec}), we propose a similar expression
for the semi-relativistic $\Delta$ current
\begin{equation}
\nj^{\Delta}_{SR}(\np,\nh)
=\frac{1}{\sqrt{1+\tau}}\nj^{\Delta}_{NR}(\np,\nh)\ ,
\label{srdelta}
\end{equation}
where $\nj^{\Delta}_{NR}(\np,\nh)$ is the standard non-relativistic
$\Delta$ current commonly considered in the
literature~\cite{Hok73,Ris89}, usually obtained from NR sets of
$\gamma N \Delta$ and $\pi N \Delta$ Lagrangians. 
In appendix B we derive its expression by performing a
non-relativistic reduction of the relativistic $\Delta$ current
considered in this work.

\begin{figure}[t]
\begin{center}
\leavevmode 
\def\epsfsize#1#2{0.9#1}
\epsfbox[50 520 478 725]{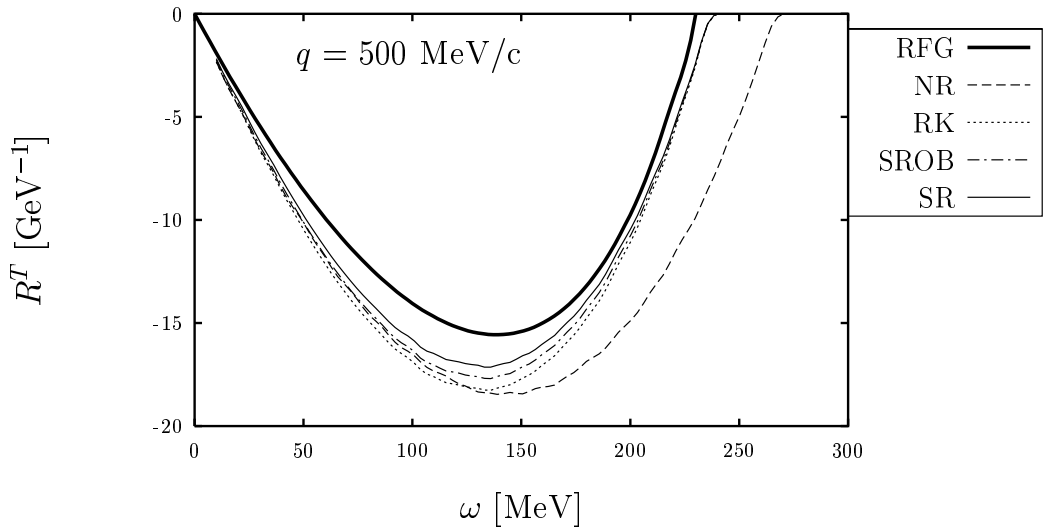}
\end{center}
\caption{MEC($\Delta$) transverse response for $q=1$ GeV/c. The result
corresponding to the fully relativistic Fermi gas (RFG) is compared
with the non-relativistic (NR), including relativistic kinematics
(RK), semi-relativistic approach for the one-body current (SROB) and
for the $\Delta$-current as well (SR) (see text for details).  }
\end{figure}

To illustrate clearly the impact of relativity we show in fig.~8 the
MEC($\Delta$) transverse response corresponding to the fully
relativistic calculation (solid lines) and to the
standard NR reduction (dashed lines) using the Fermi gas model of
ref.~\cite{Ama94}. In both cases no $\pi NN$ form factor has been
used and the static limits of the pion and $\Delta$ propagators
have been assumed.  We observe that the two calculations give the same
results for small density and momentum transfer and start to differ as
$q$ and $k_F$ increase.  In particular the results of fig.~8 provide
a test of the multi-dimensional numerical integration
procedure used in this work, since the $\Delta$ integrals over $\nk$
in the non-relativistic calculation of~\cite{Ama94} are analytical.
Moreover the agreement of both results for low
momenta also represents a test of the applicability of the
non-relativistic $\Delta$ current in this kinematic domain. 
This result is not obvious {\em a priori} due to
the somewhat coarse approximations assumed in the derivation of the
NR current (see Appendix B).

The shrinking of the response domain in fig.~8 with increasing $q$
arises from the relativistic kinematics in the energy-conserving delta
function appearing in the responses.  As in past work this 
effect can be accounted for
approximately by the replacement $\lambda \rightarrow \lambda
(1+\lambda)$, with $\lambda=\omega/2m_N$, in the
non-relativistic calculation.

This is shown in fig.~9, where we select the case $q=500$ MeV/c and
compare the NR calculation (dashed line) with the same calculation but
using relativistic kinematics (RK) (dotted line).  Apart from this
effect, the enhancement of the non-relativistic calculations with
respect to the relativistic one is a consequence of the reduction of
the single-nucleon and $\Delta$ current operators.  The
semi-relativistic approach for the single-nucleon current (SROB)
(dot-dashed) improves the agreement with the fully relativistic
result. Similarly, the response obtained by using the SR current
(\ref{srdelta}) 
(thin solid line) gets a little bit closer to the relativistic result,
although a difference between them still exists. Better agreement
between the exact and the relativized models for the MEC($\Delta$)
responses are obtained  in the limit $k_F\rightarrow 0$ in the
quasielastic peak region.

\section{Conclusions}

In this paper we have studied the role of the MEC($\Delta$) two-body
current in 1p-1h quasielastic electron scattering. A fully relativistic,
gauge invariant, analysis has been performed in the context of the
RFG.  The most general Lagrangian has been considered, allowing for
different off-shell parameters in each one of the vertices, and the
roles played by its various pieces have been analyzed. The results have been
compared with various non-relativistic reductions and the behavior of
the $\Delta$ contribution with respect to the scaling phenomenon has
been also explored.

In summary, our results show that:

\begin{enumerate}
\item
The $\Delta$ contribution is only important in the transverse channel, where 
it represents the dominant correction to the free Fermi gas, at least for 
$q<1$--$2$ GeV/c.

\item
In the transverse channel the main contribution comes from the first term of 
the Lagrangian (Peccei). This is strictly valid for not too high values of $q$.

\item
In the longitudinal channel the global $\Delta$ contribution is negligible.
In this case, however, the first and second terms of the Lagrangian tend to 
cancel.

\item
The transverse response is quite insensitive to the different sets of
off-shell parameters used, whereas the longitudinal one shows a much
more pronounced sensitivity to them.

\item
The MEC($\Delta$) contributions by themselves clearly break both 
kinds of scaling behaviour. When combined with the RFG the total
scaling breaking is on the edge of being an effect that should be
seen in the high quality high-$q$ data; however, it is known that
contributions from 2p-2h MEC tend to cancel the 1p-1h MEC effects
studied here, and thus the net scale breaking is in reasonable accord
with the data.

\item
The effects of relativity are already sizable at $q=500$ MeV/c and they can
only be accounted for approximately by using relativistic kinematics and
correcting the operators with a semi-relativistic prescription similar
to the one found in the OB current and MEC operators.

\end{enumerate}

Finally, with the present relativistic model we  have completed 
the gauge-invariant  MEC model of ref.~\cite{Ama02b} to a fully
consistent, gauge invariant approach to the 
electro-nuclear responses in the one-nucleon emission channel, 
including all the diagrams with one-pion exchange and intermediate
excitations of $\Delta$ isobar, which can be applied for high-$|Q^2|$ 
values in the region of the quasielastic peak.

\appendix

\section{$\Delta$-propagator and contact invariance}

In this Appendix we derive the general relativistic
expression for the
$\Delta$ propagator, discussing in particular its sign and
the invariance of the theory under contact
transformations.

The current amplitudes of fig.~1 are computed using the
standard Feynman rules which are essentially based on
contraction of pairs of fields or propagators.  The
$\Delta$-propagator in coordinate space is then derived, in analogy
with the Feynman propagator for spin-1/2 fermions. It reads
\begin{equation}
i G_{RS}^{\alpha\beta}(X-X') = 
\langle0|T\left\{\psi^\alpha(X)\overline{\psi}^\beta(X')\right\}|0\rangle
\ ,
\label{timeord}
\end{equation}
where 
\begin{equation}
\psi^\alpha(X) 
=\sum_{\np,s_\Delta} \sqrt{\frac{m_\Delta}{V E_\np}}
\left[
c_{s_\Delta} u^\alpha(\np,s_\Delta) e^{-i P\cdot X}+
d_{s_\Delta}^{\dagger} v^\alpha(\np,s_\Delta) e^{+i P\cdot X}
\right]
\label{deltafield}
\end{equation}
is the free massive spin-3/2 Rarita-Schwinger field, resulting from the
coupling  of a spin-1/2 and a spin-1 object:
\begin{equation}
u^\alpha(\np,s_\Delta)=\sum_{\lambda,s}
\textstyle
\langle \frac{1}{2} s 1 \lambda|\frac{3}{2} s_\Delta \rangle 
e^\alpha(\np,\lambda) u(\np,s) \ ,
\label{ualpha}
\end{equation}
$u(\np,s)$ being a Dirac spinor of mass $m_\Delta$
and $e^\alpha(\np,\lambda)$ the basis vectors. 

The Rarita-Schwinger propagator in momentum space is the 4-dimensional
Fourier transform of $G_{RS}(X)$ 
\be
G_{RS}^{\alpha\beta}(X) =
\int\frac{d^4 P}{(2\pi)^4} G_{RS}^{\alpha\beta}(P) 
e^{-iP\cdot X}
\label{iSRS}
\ee
and it reads
\be
G_{RS}^{\alpha\beta}(P)=-\frac{ \Pbar+m_\Delta}{P^2-m_\Delta^2}
\left[ g^{\alpha\beta} - \frac{1}{3} \gamma^\alpha\gamma^\beta - 
\frac{2}{3} \frac{P^\alpha P^\beta}{m_\Delta^2} - 
\frac{1}{3} \frac{\gamma^\alpha P^\beta - \gamma^\beta P^\alpha}{m_\Delta} 
\right]
\ .
\label{GRS}
\ee Note that the minus sign in eq.~(\ref{GRS}), directly arising from
the definition in eq.~(\ref{timeord}), is opposite to the one in
refs.~\cite{Ben89,Dav91,Pas95}, whereas it agrees with
ref.~\cite{Amo01}.  A simple check of the sign can be done by
evaluating the right-hand side of eq.~(\ref{timeord}) for
$t>t'$. Inserting the field expansion of eq.~(\ref{deltafield}) in
eq.~(\ref{timeord}) we obtain
\begin{equation}
iG_{RS}^{\alpha\beta}(X-X')
= \int\frac{d^3p}{(2\pi)^3} \frac{m_\Delta}{E}
\sum_{s_\Delta}
  u^\alpha(\np,s_\Delta)\overline{u}^\beta(\np,s_\Delta)
  e^{-iP\cdot(X-X')}.
\end{equation}
The spin sum inside the integral reads
\begin{eqnarray}
\lefteqn{
          \sum_{s_\Delta}
          u^\alpha(\np,s_\Delta)\overline{u}^\beta(\np,s_\Delta)
        }
\nonumber\\
&=&
\sum_{s_\Delta}\sum_{\lambda s}\sum_{\lambda' s'}
\textstyle
\langle \frac{1}{2} s 1 \lambda|\frac{3}{2} s_\Delta \rangle 
\langle \frac{1}{2} s' 1 \lambda'|\frac{3}{2} s_\Delta \rangle 
e^\alpha(\np,\lambda) 
e^\beta(\np,\lambda')^*
 u(\np,s)  \overline{u}(\np,s') \ .
\end{eqnarray}
This sum is particularly simple 
in the static limit $\np=0$ and for the 
components $\alpha=\beta=3$: in this case 
$e^\alpha(\np,\lambda)$ and  
$e^\beta(\np,\lambda')$ vanish unless $\lambda=\lambda'=0$. Moreover the
Clebsch-Gordan coefficients select  $s=s'=s_\Delta$, yielding
\begin{eqnarray}
\sum_{s_\Delta}
 u^3(0,s_\Delta)\overline{u}^3(0,s_\Delta)
&=&
\sum_{s}        
\textstyle
\langle \frac{1}{2} s 1 0|\frac{3}{2} s \rangle^2 \
 u(0,s)  \overline{u}(0,s) 
\nonumber \\
&=&
\frac23 \sum_s  
 u(0,s)  \overline{u}(0,s) =
\frac23 \frac{m_\Delta+\Pbar}{2m_\Delta}\, ,
\label{projector}
\end{eqnarray}
where
we have used 
$\langle \frac12\frac12 1 0 |\frac32\frac12\rangle=
\langle \frac12 -\frac12 1 0 |\frac32 -\frac12\rangle=\sqrt{\frac23}$,
and 
where the positive-energy projector for spin 1/2 and mass $m_\Delta$
is meant to be evaluated at $\np=0$.
The same result is obtained from the RS propagator
in eq.~(\ref{GRS}) setting $\np=0$ and $\alpha=\beta=3$,
performing the contour integral over $p_0$ in eq.~(\ref{iSRS}):
\begin{equation}
i\int \frac{d p_0}{2\pi} 
 \frac{ \Pbar+m_\Delta}{P^2-m_\Delta^2+i\epsilon}
 e^{-ip_0 t}
=
 \frac{ \Pbar+m_\Delta}{2E_{\np}} 
\end{equation}
and using
\begin{equation}
G^{33}_{RS}(P)=
-\frac{\Pbar+m_\Delta}{P^2-m_\Delta^2}
\left[ g^{33}-\frac13(\gamma^3)^2\right]
=
\frac23\frac{\Pbar+m_\Delta}{P^2-m_\Delta^2}.
\end{equation}

Let us now discuss the so-called ``contact-invariance'', induced by the
operator
\begin{equation}
O^{\mu}{}_{\nu}(b)=g^{\mu}{}_\nu+\frac{b}{2} S^{\mu}{}_{\nu}\ ,
\end{equation}
where $b$ is an arbitrary parameter and 
$S^{\mu}{}_{\nu}\equiv\gamma^\mu\gamma_\nu$ is the generator of the
transformation, fulfilling the identity $S^2=4S$.

A contact transformation of the RS field is defined by
\begin{equation}
\psi'{}^\alpha=O^\alpha{}_\beta(b) \psi^\beta\ .
\label{contact}
\end{equation}
Some useful properties of the contact operators are:
\begin{eqnarray}
& \mbox{i)}  & O(a)O(b)=O(a+b+2a b)
\label{prop1}
\\
& \mbox{ii)} & O^{-1}(b)= O\left(-\frac{b}{1+2b}\right) 
\label{prop2}
\\
& \mbox{iii)} & O(a)O^{-1}(b)= O\left(\frac{a-b}{1+2b}\right) \ .
\label{prop3}
\end{eqnarray}

The contact invariance follows from the fact that the $A$-dependent free
Lagrangian of the $\frac32$-field becomes independent of the parameter
$A$, after a redefinition of the RS field~\cite{Pas95}:
\begin{equation}
\Delta^\alpha = O^\alpha{}_\beta(A)\psi^\beta \ .
\end{equation} 
Therefore, in terms of the contact transformation eq.~(\ref{contact}) we
can write the new field in matrix form as:
\begin{equation}
\Delta=O(A)O^{-1}(b)\psi'
\end{equation}
and using eq.~(\ref{prop3}) we obtain that
contact invariance is fulfilled if the parameter $A$ changes to
$A'=(A-b)/(1+2b)$.

We now derive the general form of the $\Delta$-propagator 
$G(A)$ for an arbitrary parameter $A$ in terms of the usual RS propagator
in eq.~(\ref{GRS}). From the definition in eq.~(\ref{timeord})
 (for simplicity we omit the $X,X'$ variables) we get
\begin{eqnarray}
i G_{\alpha\beta}(A)  
&=& 
\langle0|T \left\{ O^{-1}_{\alpha\mu}(A)\Delta^\mu\overline{\Delta}^\nu
   O^{-1}_{\nu\beta}(A) \right\}|0\rangle
\nonumber\\
&=&
 O^{-1}_{\alpha\mu}(A)
  \langle0|T\{  \Delta^\mu  \overline{\Delta}^\nu  \}|0\rangle 
  O^{-1}_{\nu\beta}(A)
\nonumber\\
&=&
O^{-1}_{\alpha\mu}(A) iG^{\mu\nu}(0)O^{-1}_{\nu\beta}(A) \, ,
\end{eqnarray}
where $G^{\mu\nu}(0)$ is the propagator for $A=0$, obtained in terms of the
free Lagrangian for the $\Delta^\alpha$ field. 
The above equation can be written in matrix form introducing the 
$4\times4$ matrix  of the propagator $G(A)\equiv G^\mu{}_\nu(A)$
\begin{equation}
G(A)=O^{-1}(A) G(0) O^{-1}(A)\ .
\end{equation}
The RS propagator in eq.~(\ref{GRS}) corresponds to $A=-1$, which also
can be related with the $A=0$ case
\begin{equation}
G(-1)=O^{-1}(-1) G(0) O^{-1}(-1)\ .
\end{equation}
Using the two above equations we can write
\begin{eqnarray}
G(A) 
&=& 
O^{-1}(A) O(-1) G(-1) O(-1) O^{-1}(A) 
\nonumber\\
&=&
 O(B) G(-1) O(B)\, ,
\label{Ginv}
\end{eqnarray}
where 
\begin{equation}
B=-\frac{A+1}{1+2A}
\label{B}
\end{equation}
and use has been made of eq.~(\ref{prop3}).

From eq.~(\ref{Ginv}) it is immediate to show that 
the
combination $X(A)\equiv\Theta(z,A) G(A) \Theta(z',A)$, appearing in the 
MEC($\Delta$) current,
is independent of $A$. We note that the operator in eq.~(\ref{Theta})
can be expressed, by means of eq.~(\ref{prop1}),
as the product of two contact operators
\begin{equation}
\Theta(z,A)=O(2z) O(A)=O(A)O(2z) \, .
\end{equation}
Hence
\begin{equation}
X(A)=O(2z)O(A)G(A)O(A)O(2z')=O(2z)G(0)O(2z)= X(0)\ .
\end{equation}
To obtain the explicit form of $G(A)$, 
we note that the 
Rarita-Schwinger propagator  $G(-1)$, given by 
eq.~(\ref{GRS}),  can be more conveniently written
as
\be
G(-1)=-\frac{ \Pbar+m_\Delta}{P^2-m_\Delta^2}
\left[1\!\!1 - \frac{1}{3} S-\frac{2}{3m_\Delta^2} U-\frac{1}{3 m_\Delta}
(V-W)
\right]\ ,
\label{RS1}
\ee
having introduced the matrices
\be
U^\mu{}_\nu \equiv  P^\mu P_\nu\ ,\ 
V^\mu{}_\nu \equiv  \gamma^\mu P_\nu\ ,\ 
W^\mu{}_\nu \equiv  P^\mu \gamma_\nu\ .
\ee
From eq.~(\ref{Ginv}) it follows that
\be
G(A)= G(-1)+\frac{B}{2}\left[ S G(-1)+G(-1) S \right]
+\frac{B^2}{4} SG(-1)S
=G_{RS}+G_A\ .
\ee
After straightforward $\gamma$-matrix algebra the following useful relations
can be deduced:
\ba
SG(-1)+G(-1)S &=& \frac{2}{3 m_\Delta^2}(V+W-m_\Delta S)
\\
S G(-1) S &=& \frac{2}{3 m_\Delta^2}(2 W -\Pbar S-2 m_\Delta S)
\ea
which finally yield
\be
G_A=
\frac{1}{3m_\Delta^2} B
\left[ V + W - m_\Delta S
+ \frac{1}{2} B (2 W-\Pbar S-2 m_\Delta S)\right]\ .
\ee
This, exploiting eq.~(\ref{B}), coincides with eq.~(\ref{Aprop}).


\section{Non-relativistic reduction of the $\Delta$-current}


In this appendix we derive in detail the 
non-relativistic reduction of the $\Delta$-exchange
current corresponding to the Peccei-like $\gamma N \Delta$ vertex.

Let us consider the $\gamma N\Delta$ Lagrangian as given by eq.~(\ref{lag1}).
The corresponding two-body current  can be written as follows
\begin{eqnarray}
j^\mu_\Delta(\np'_1,\np'_2,\np_1,\np_2) 
&=&  \frac{G_1}{2m_N}\frac{f_{\pi N\Delta}f}{m_\pi^2}
G_{\pi}(K_2)
\left[A^\mu T_aT_3^{\dagger} + B^\mu  T_3T_a^{\dagger} \right]
   \overline{u}(\np'_2)\gamma_5\Kbar_2\tau^a u(\np_2)
\nonumber\\ 
&+& (1\iff 2)\, ,
\label{appB1}
\end{eqnarray}
where we have introduced
\begin{eqnarray}
A^\mu &\equiv & 
\overline{u}(\np'_1)K_{2\alpha}
\Theta^{\alpha\beta}
G^{RS}_{\beta\rho}(P_1+Q)Q_\nu
\left(\Theta^{\rho\mu}\gamma^\nu-\Theta^{\rho\nu}\gamma^\mu\right)
\gamma_5  u(\np_1) \label{amu}
\\
B^\mu &\equiv &
\overline{u}(\np'_1)
K_{2\alpha}
\gamma_5 
Q_\nu
\left(\gamma^\nu\Theta^{\mu\rho}-\gamma^\mu\Theta^{\nu\rho}\right)
G^{RS}_{\rho\beta}(P'_1-Q)
\Theta^{\beta\alpha}
u(\np_1) \, . \label{bmu}
\end{eqnarray}
To compare with the Peccei vertex
the tensor $\Theta^{\alpha\beta}$ in eq.~(\ref{Theta}) should be
taken for $z=-1/4$ and the $\Delta$ propagator is the Rarita-Schwinger
expression in eq.~(\ref{Dprop}).

In what follows we invoke the static limit as usually considered in
standard non-relativistic calculations. In this case the spin dependence
in the tensor $\Theta^{\alpha\beta}$ is neglected, i.e., 
$\Theta^{\alpha\beta}=g^{\alpha\beta}$. Moreover, the pion propagator
simply reduces to
\begin{equation}
G_\pi(K) = -\frac{1}{\nk^2+m_\pi^2}
\end{equation}
while the $\Delta$ propagator only contributes for space indices becoming
\begin{equation}
G^\Delta_{ij}(P) =
\frac{1}{m_N-m_\Delta}\left(\delta_{ij}+\frac13\gamma_i\gamma_j\right),
\kern 1cm i,j=1,2,3 \, .
\end{equation}
Assuming $K_2^\alpha\sim (0,\nk_2)$ and
$Q^\mu\sim (0,\nq)$ (valid in the static limit), the space components 
of the four-vectors $A^\mu$ and $B^\mu$ in 
eqs.~(\ref{amu},\ref{bmu}) can be written
in terms of space components only 
\begin{eqnarray}
A^i
&\simeq&
-\frac{1}{m_N-m_\Delta}
\overline{u}(\np'_1)
k_2^k
\left(\delta_{kl}+\frac13\gamma_k\gamma_l\right)
q_j(-\delta^{li}\gamma^j+\delta^{lj}\gamma^i)\gamma_5 u(\np_1)
\\
B^i
&\simeq&
-\frac{1}{m_N-m_\Delta}
\overline{u}(\np'_1)
k_2^k
q_j(-\delta^{li}\gamma^j+\delta^{lj}\gamma^i)\gamma_5 
\left(\delta_{kl}+\frac13\gamma_l\gamma_k\right)
u(\np_1)\ .
\end{eqnarray}
Taking now the positive energy components of the gamma matrices,
\begin{equation}
\gamma_k\gamma_l\rightarrow -\sigma_k\sigma_l,
\kern 2em
\gamma_i\gamma_5 \rightarrow \sigma_i,
\kern 2em
\delta_{kl}+\frac13\gamma_l\gamma_k \rightarrow 
\frac23\delta_{kl}-\frac{i}{3}\epsilon_{klm}\sigma_m
\end{equation}
we can write $A^i$, $B^i$ in terms of matrix elements between 
 the Pauli spinors, $\chi_1$, as follows
\begin{eqnarray}
A^i &=&
\frac{1}{m_N-m_\Delta}
\chi'_1{}^{\dagger} \overline{A^i} \chi_1
\\
B^i &=&
\frac{1}{m_N-m_\Delta}\chi'_1{}^{\dagger} \overline{B^i} \chi_1 \, ,
\end{eqnarray}
where $\overline{A^i}$ and $\overline{B^i}$ are the non-relativistic
reduced operators given by
\begin{eqnarray}
\overline{A^i} 
&=&
\frac23 k_2^i q_j \sigma_j
-\frac23 k_2^j q_j \sigma_i
-\frac{i}{3}\epsilon_{kim} k_{2k} q_j\sigma_m \sigma_j
+\frac{i}{3}\epsilon_{kjm} k_{2k} q_j\sigma_m \sigma_i
\\
\overline{B^i}
&=& 
-\frac23 k_{2i} q_j \sigma_j 
+\frac23 k_{2j} q_j \sigma_i 
+\frac{i}{3}\epsilon_{ikm} k_{2k}  q_j \sigma_j \sigma_m
-\frac{i}{3}\epsilon_{jkm} k_{2k}  q_j \sigma_i \sigma_m \, .
\end{eqnarray}
After some algebra involving vector relations and properties of the 
Pauli matrices, 
the reduced non-relativistic $A$ and $B$ terms in their vector form result
\begin{eqnarray}
\overline{\nA} &=& 
\frac13\nq\times(\nk_2\times\nsigma)
-\frac{2}{3}i\nq\times\nk_2
\\
\overline{\nB} &=& 
-\frac13\nq\times(\nk_2\times\nsigma)
-\frac{2}{3}i\nq\times\nk_2 \, .
\end{eqnarray} 

Moreover, the matrix element of the second nucleon in the static limit reduces to
\begin{equation}
\overline{u}(\np'_2)\gamma_5\Kbar_2 u(\np_2)
\longrightarrow
\overline{u}(p'_2)\nk_2\cdot\ngamma \gamma_5 u(p_2)
\longrightarrow
\chi'_2{}^{\dagger}
\left(\nk_2\cdot\nsigma\right) \chi_2 \, .
\end{equation}
Using the above results, the $i$th component of the current 
can be written as
\begin{equation}
J^i_\Delta = \chi'_1{}^{\dagger}\chi'_2{}^{\dagger}\overline{J^i}_\Delta\chi_1\chi_2 \, ,
\end{equation}
where, using eq.~(\ref{iso}),
\begin{eqnarray}
\overline{J^i}_\Delta
&\simeq&
i\frac29 
\frac{G_1}{2m_N}\frac{f_{\pi N\Delta}}{m_\pi}\frac{f}{m_\pi}
\frac{\nk_2\cdot\nsigma^{(2)}}{m_\pi^2+\nk_2^2}\frac{1}{m_\Delta-m_N}
\left\{
        4\tau_3^{(2)}\nk_2
      -\left[\tauvec^{(1)}\times\tauvec^{(2)}\right]_z
      \nsigma^{(1)}\times\nk_2
\right\}\times\nq
\nonumber\\
&& 
\mbox{}+(1\longrightarrow 2) \, .
\nonumber\\
\end{eqnarray}
This form coincides with the usual non-relativistic
$\Delta$ current used in the literature~\cite{Ris89}.

\section*{Acknowledgments}
This work was partially supported by funds provided by DGI (Spain) and
FEDER funds, under Contracts Nos BFM2002-03218, BFM2002-03315 and
FPA2002-04181-C04-04 and by the Junta de Andaluc\'{\i}a and by the
INFN-CICYT exchange.  M.B.B. acknowledges financial support from MEC
(Spain) for a sabbatical stay at University of Sevilla
(ref. SAB2001-0025). T.W.D. was supported by funds provided by the
U.S. Department of Energy under cooperative research agreement
No. DE-FC02-94ER40818.


\end{document}